\documentclass[useAMS,usenatbib,usegraphicx]{mn2e}
\usepackage{color}
\usepackage{wasysym}
\usepackage{url}
\usepackage{multirow}
\usepackage{times}
\usepackage[usenames,dvipsnames]{xcolor}
\usepackage{hyperref}

\newcommand{\nuc}[2]{\ensuremath{\mathrm{^{#1}#2}}}
\newcommand{\ions}[2]{#1\,{\sc #2}}

\newcommand{\msun}{\ensuremath{\mathrm{M}_\odot}}

\def\lesssim{\mathrel{\hbox{\rlap{\hbox{\lower4pt\hbox{$\sim$}}}\hbox{$<$}}}}

\def\gtrsim{\mathrel{\hbox{\rlap{\hbox{\lower4pt\hbox{$\sim$}}}\hbox{$>$}}}}

\def\aj{AJ}%
\def\araa{ARA\&A}%
\def\apj{ApJ}%
\def\apjl{ApJL}%
\def\apjs{ApJS}%
\def\apss{Ap\&SS}%
\def\aap{A\&A}%
\def\mnras{MNRAS}%
\def\pasp{PASP}%
\def\nat{Nature}%
\title[Modelling iPTF14atg] {%
  The peculiar Type Ia supernova iPTF14atg: 
  Chandrasekhar-mass explosion or violent merger? }

\author[M.~Kromer~et~al.]
{M.~Kromer,$^{1}$\thanks{E-mail: markus.kromer@astro.su.se}
  C.~Fremling,$^{1}$
  R.~Pakmor,$^{2}$
  S.~Taubenberger,$^{3,4}$
  R.~Amanullah,$^{5}$
  \newauthor
  S.~B.~Cenko,$^{6}$
  C.~Fransson,$^{1}$
  A.~Goobar,$^{5}$
  G.~Leloudas,$^{7,8}$
  F.~Taddia,$^{1}$
  F.~K.~R\"{o}pke,$^{2,9}$
  \newauthor
  I.~R.~Seitenzahl,$^{10,11}$
  S.~A.~Sim$^{11,12}$
  and J.~Sollerman$^{1}$\\
  $^{1}$The Oskar Klein Centre \& Department of Astronomy,
       Stockholm University, AlbaNova, SE-106 91 Stockholm, Sweden\\
  $^{2}$Heidelberger Institut f\"{u}r Theoretische Studien, 
        Schloss-Wolfsbrunnenweg 35, D-69118 Heidelberg, Germany\\
  $^{3}$European Southern Observatory, 
       Karl-Schwarzschild-Str. 2, D-85748 Garching bei M{\"u}nchen, Germany\\  
  $^{4}$Max-Planck-Institut f{\"u}r Astrophysik, 
       Karl-Schwarzschild-Str. 1, D-85748 Garching bei M{\"u}nchen, Germany\\
  $^{5}$The Oskar Klein Centre \& Department of Physics,
       Stockholm University, AlbaNova, SE-106 91 Stockholm, Sweden\\
  $^{6}$Astrophysics Science Division, NASA Goddard Space Flight
       Center, Mail Code 661, Greenbelt, MD 20771, USA\\
  $^{7}$Department of Particle Physics and Astrophysics, Weizmann
       Institute of Science, Rehovot 7610001, Israel\\
  $^{8}$Dark Cosmology Centre, Niels Bohr Institute, University of
       Copenhagen, Juliane Maries vej 30, 2100 Copenhagen, Denmark\\
  $^{9}$Zentrum f{\"u}r Astronomie der Universit{\"a}t Heidelberg, 
       Institut f{\"u}r Theoretische Astrophysik, Philosophenweg 12, 
       D-69120 Heidelberg, Germany\\
  $^{10}$Research School of Astronomy and Astrophysics, 
       Australian National University, Canberra, ACT 2611, Australia\\
  $^{11}$ARC Centre of Excellence for All-Sky Astrophysics (CAASTRO)\\
  $^{12}$Astrophysics Research Centre, School of Mathematics and Physics, 
       Queen's University Belfast, Belfast BT7 1NN, UK\\
}

\begin{document}

\date{Accepted, 20 April 2016. Received, 12 April 2016; in original form, 19
February 2016}
\pagerange{\pageref{firstpage}--\pageref{lastpage}} \pubyear{2016}

\maketitle

\label{firstpage}

\begin{abstract}
  iPTF14atg, a subluminous peculiar Type Ia supernova (SN~Ia) similar
  to SN~2002es, is the first SN~Ia for which a strong UV flash was
  observed in the early-time light curves. This has been interpreted
  as evidence for a single-degenerate (SD) progenitor system where
  such a signal is expected from interactions between the SN ejecta
  and the non-degenerate companion star. Here, we compare synthetic
  observables of multi-dimensional state-of-the-art explosion models
  for different progenitor scenarios to the light curves and spectra
  of iPTF14atg. From our models, we have difficulties explaining the
  spectral evolution of iPTF14atg within the SD progenitor channel.
  In contrast, we find that a violent merger of two carbon-oxygen
  white dwarfs with 0.9 and 0.76\,\msun, respectively, provides an
  excellent match to the spectral evolution of iPTF14atg from 10\,d
  before to several weeks after maximum light. Our merger model does
  not naturally explain the initial UV flash of iPTF14atg. We discuss
  several possibilities like interactions of the SN ejecta with the
  circum-stellar medium and surface radioactivity from a He ignited
  merger that may be able to account for the early UV emission in
  violent merger models.
\end{abstract}


\section{Introduction}

Despite the importance of Type Ia supernovae (SNe~Ia) for cosmological
distance measurements \citep{riess1998a,perlmutter1999a} and decades
of intensive work, their progenitors are still elusive. It is widely
accepted that SNe~Ia result from thermonuclear explosions in
carbon--oxygen (CO) white dwarfs (WDs) that are triggered by some kind
of interaction in a binary system \citep[see e.g.][for a
review]{hillebrandt2000a}. However, the exact nature of the binary
system is still under debate.

In the most commonly discussed scenario the CO WD accretes H from a
non-degenerate companion star and explodes due to the onset of
pycnonuclear reactions when the WD nears the Chandrasekhar-mass -- the
so called single-degenerate (SD) scenario
\citep{whelan1973a,nomoto1982a}.  Alternatives are the
double-degenerate (DD) scenario where two CO WDs merge due to
gravitational wave emission \citep{iben1984a,webbink1984a}, or double
detonations in sub-Chandrasekhar-mass He accreting CO WDs where the
companion star can either be a He WD or He-burning star
\citep{nomoto1980a,nomoto1982b,woosley1980a,woosley1984a}.

The different progenitor scenarios leave characteristic imprints on
the observational properties of SNe~Ia, which have been used to obtain
constraints on the progenitors (for a detailed review see
\citealt{maoz2014a}). Unfortunately, the results of these analyses are
not fully conclusive. On the one hand, the non-detection of companion
stars in deep pre-explosion images of SN~2011fe and SN~2014J
\citep[e.g.][]{li2011b, kelly2014a}, the lack of X-ray and radio
emission \citep[e.g.][]{chomiuk2012a, margutti2012a,
  perez-torres2014a}, the absence of H features in the late-time
spectra of SNe~Ia \citep[e.g.][]{shapee2013a, lundqvist2015a,
  maguire2016a} and the non-detection of surviving companion stars in
historic SN~Ia remnants \citep[e.g.][]{kerzendorf2009a,schaefer2012a}
favours DD progenitors. On the other hand, some SNe~Ia show signs of
circum-stellar material \citep[e.g.][]{patat2007a, dilday2012a}, which
are typically explained as the result of mass-transfer phases in SD
progenitor systems (but see \citealt{shen2013a}). Furthermore, the
chemical evolution of [Mn/Fe] requires that thermonuclear burning in
near-Chandrasekhar mass WDs significantly contributed to the solar Fe
abundance (see \citealt{seitenzahl2013b}), which is most easily
reconciled with some SNe~Ia arising from SD progenitors.

Another smoking-gun signature of SD progenitors results from the
collision of the SN ejecta with its non-degenerate companion
star. \citet{kasen2010a} has shown that this collision leads to
reheating of parts of the ejecta and strong emission in the UV and
blue bands for a few days after explosion. The strength and duration
of the UV emission depend strongly on the binary separation and the
viewing angle. Applying Kasen's models to results from binary
population synthesis calculations, \citet{liu2015a} present the
expected UV luminosity distribution for a variety of SD progenitor
systems. Previous attempts at detecting ejecta-companion interactions
in indvidual nearby SNe~Ia \citep{nugent2011a, bloom2012a, brown2012a,
  goobar2015a, olling2015a} and samples of SNe~Ia from cosmological
surveys \citep{hayden2010b, bianco2011} have been unsuccessful.
Recently, however, the detections of an early ultraviolet flash in the
peculiar SN~Ia iPTF14atg \citep{cao2015a} and an excess of blue light
in the normal Type Ia SN~2012cg \citep{marion2016a} have been
interpreted as evidence for SD progenitors of these SNe.

Here, we investigate iPTF14atg in the context of state-of-the-art
explosion models for different progenitor and explosion scenarios.
The paper is organized as follows. In Section~\ref{sec:iptf14atg} we
compile the observational properties of iPTF14atg and present a new
late-time spectrum of the SN. In Section~\ref{sec:models} we compare
the observed spectra and light curves of iPTF14atg to synthetic
observables from hydrodynamic explosion models for a SD progenitor and
a WD merger. We discuss our findings in Section~\ref{sec:discussion},
before concluding in Section~\ref{sec:conclusions}.

\section{\lowercase{i}PTF14\lowercase{atg}}
\label{sec:iptf14atg}

iPTF14atg was discovered on May 3.29, 2014 (UTC) when the intermediate
Palomar Transient Factory (iPTF, \citealt{law2009a}) detected a new
point source in the early-type galaxy IC~831 \citep{cao2015a}. The
source was not present in observations on May 2.29, suggesting that
the SN was caught very early with a likely explosion date between May
2.29 and 3.29. Here, we follow \citet{cao2015a} and assume May 3.0 for
the time of explosion.

In the following weeks, the iPTF collaboration performed comprehensive
observations of the new transient and obtained multi-band photometry
with the Ultraviolet/Optical Telescope (UVOT) onboard the
\textit{Swift} space observatory and from various ground based
observatories. They also followed the spectral evolution of iPTF14atg
from early epochs to about 3 months after the explosion. The resulting
observables and details on the data reduction are presented in
\citet{cao2015a}. For our study we have calibrated the spectral time
sequence with respect to the PTF P48 R-band photometry.

From their observations, \citet{cao2015a} classified iPTF14atg as a
slowly-evolving subluminous SN~Ia with an absolute magnitude of
$-17.9$ in the $B$ band and a low expansion velocity. Comparing
iPTF14atg to other subluminous SNe~Ia they suggest it belongs to a
class of objects similar to SN~2002es \citep{ganeshalingam2012a}.
These objects show similar spectral features to 1991bg-like SNe, but
significantly slower evolution and lower expansion velocities. Other
known objects of this class include PTF10ops \citep{maguire2011a} and
SN~2010lp \citep{kromer2013b}. A comparison to Type Iax SNe
\citep{foley2013b} was also attempted but found to be less favourable.

The extensive data set of iPTF14atg clearly makes it the best observed
member of the 2002es-like class. What makes it really special
is a high flux in the \textit{Swift} UV bands at the earliest epoch
(4\,d past explosion), which is followed by a steep decline lasting
about 1\,d before the UV flux rises again (fig.~1 of
\citealt{cao2015a}). Such an initial UV pulse has not been observed in
any other SN~Ia before. Given that no other 2002es-like SNe have been
observed at comparably early epochs, it is not clear whether this is a
unique feature of iPTF14atg or typical for this class of objects.

\citet{cao2015a} discuss several possibilities to explain the initial
UV pulse, including a layer of radioactive isotopes in the outer
ejecta, shock interaction with the circum-stellar medium and
interactions between the SN ejecta and a non-degenerate companion
star. They conclude that companion interaction is the only viable
scenario, claiming a SD progenitor for iPTF14atg.

\subsection{Nebular spectrum}

We obtained an optical spectrum of iPTF14atg on December 21, 2014
(corresponding to 232\,d past explosion) with the Deep Imaging
Multi-Object Spectrograph (DEIMOS; \citealt{faber2003a}) mounted on
the 10\,m Keck II telescope.  The instrument was set up with the
600\,lines\,mm$^{-1}$ grating, providing spectral coverage over the
region $\lambda = 4500$--9500\,\AA\ with a spectral resolution of
3.5\,\AA.  The spectrum was optimally extracted \citep{horne1986a},
and the rectification and sky subtraction were performed following the
procedure described by \citet{kelson2003a}.  The slit was oriented at
the parallactic angle to minimize losses due to atmospheric dispersion
\citep{filipenko1982a}.

\begin{figure}
  \centering
  \includegraphics[width=0.9\linewidth]{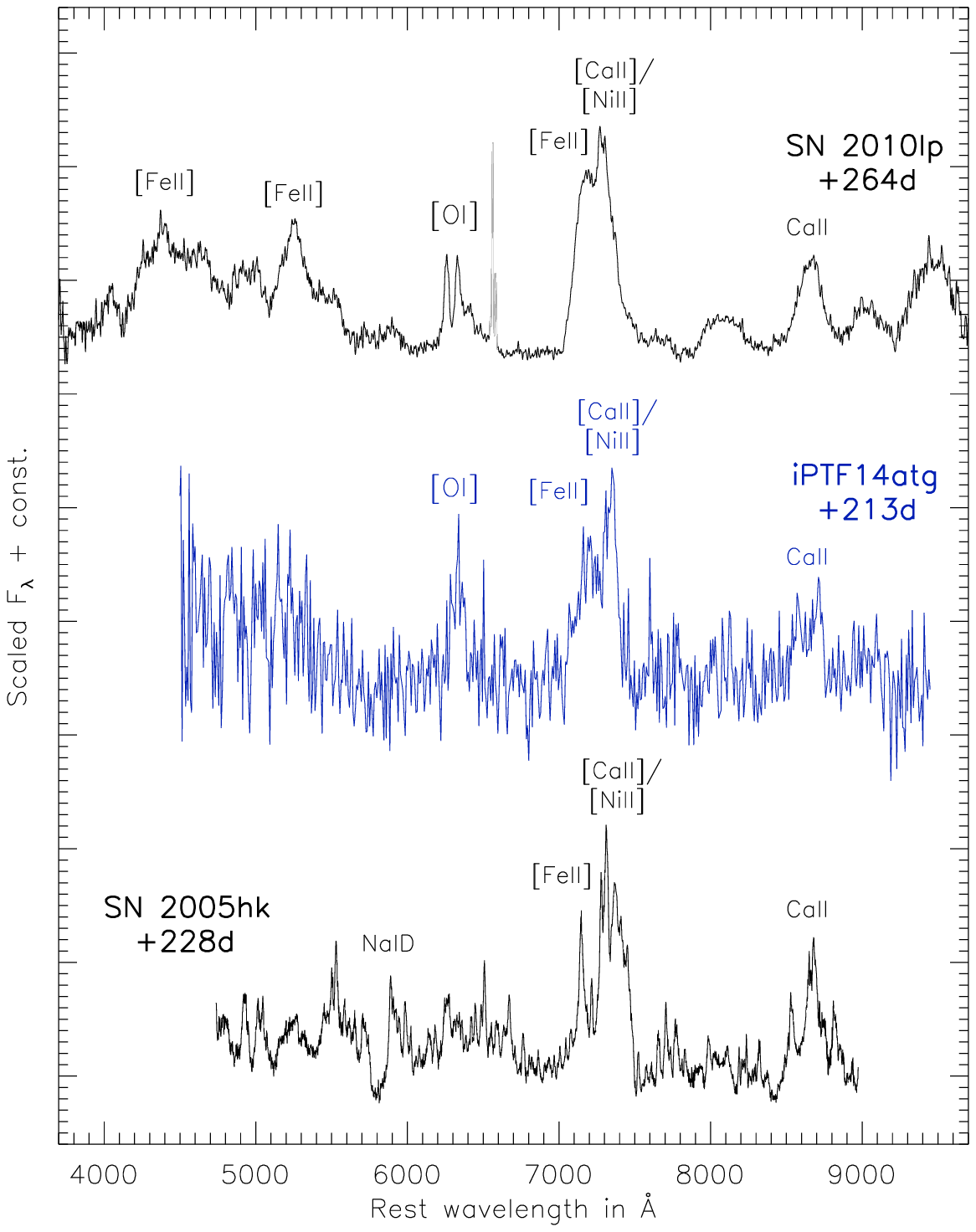}
  \caption{The nebular Keck-DEIMOS spectrum of iPTF14atg taken at
    213\,d past $B$-band maximum (blue, the spectrum was re-binned to
    a wavelength resolution of 8\,\AA\ to increase the signal-to-noise
    ratio). Note the emission feature at 6300\,\AA. For comparison
    late-time spectra of the 2002es-like SN~2010lp
    \citep{taubenberger2013b} and the Type Iax SN~2005hk
    \citep{sahu2008a} are shown at +264\,d and +228\,d,
    respectively. Narrow H$\alpha$ emission in the spectrum of
    SN~2010lp, originating from a nearby \ions{H}{ii} region, has been
    greyed out to highlight the intrinsic SN features.}
  \label{fig:nebular}
\end{figure}

This late-time spectrum, presented in Figure~\ref{fig:nebular},
clearly shows forbidden emission lines of Fe, which are the hallmark
feature of SNe~Ia in the nebular phase. Specifically, we find
pseudo-continuous [\ions{Fe}{ii}] emission blueward of
$\sim$5500\,\AA\ and a strong emission feature at 7000-7500\,\AA,
which is typically attributed to [\ions{Fe}{ii}] $\lambda7155$ and
[\ions{Ca}{ii}] $\lambda\lambda7291, 7323$.  This is very similar to
the behaviour of subluminous 1991bg-like SNe~Ia \citep{mazzali1997a,
  mazzali2012a}.

In addition, iPTF14atg shows a pronounced emission feature at
$\sim$6300\,\AA, which is not observed in 1991bg-like SNe~Ia. However,
a similar feature was observed in the 2002es-like SN~2010lp by
\citet{taubenberger2013b}, thus corroborating the classification of
iPTF14atg as a 2002es-like SN by \citet{cao2015a}. The nature of this
6300\,\AA\ feature has not yet been investigated with detailed
radiative transfer models. \citet{taubenberger2013b} attribute it to
[\ions{O}{i}] $\lambda\lambda6300,6364$ emission in analogy to the
nebular spectra of Type Ib/c SNe \citep[e.g.][]{maeda2008a,
  taubenberger2009a}. 

This would require a significant fraction of O in the central parts of
the explosion ejecta. While pure detonations and delayed detonations
of Chandrasekhar-mass WDs do not leave sufficient O in the central
ejecta \citep[e.g.][]{seitenzahl2013a}, turbulent burning in
Chandrasekhar-mass deflagrations distributes O all over the ejecta
\citep[e.g.][]{roepke2007c, ma2013a}. It has been shown that such
deflagration models indeed lead to [\ions{O}{i}] emission at late times
\citep{kozma2005a}.  However, due to the mixing, these features are
fairly broad. Therefore, \citet{taubenberger2013b} discard this model
for the narrow [\ions{O}{i}] features seen in SN~2010lp. Instead, they
favour a violent merger like that by \citet{pakmor2012a}, where
unburned O from the secondary WD is present at low velocities in the
central ejecta but no strong mixing.

A connection of iPTF14atg to Type Iax SNe, which was discussed but
already discarded by \citet{cao2015a}, seems also unlikely from a
comparison of the late-time spectra. The proto-typical Type Iax
SN~2005hk does not show an emission feature at $\sim$6300\,\AA.
Instead, late-time spectra of Type Iax SNe are still dominated by
low-velocity P-Cygni profiles of permitted Fe lines
\citep{jha2006b}. Given the poor S/N of our late-time spectrum we can
unfortunately not test this for iPTF14atg.

\begin{figure}
 \centering
 \includegraphics[width=0.9\linewidth]{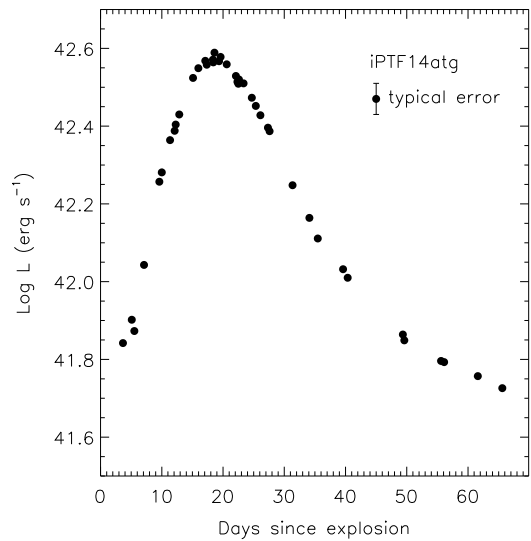}
 \caption{Pseudo-bolometric light curve of iPTF14atg.}
 \label{fig:bolo}
\end{figure}

\subsection{Bolometric light curve}

Figure~\ref{fig:bolo} shows a pseudo-bolometric light curve of
iPTF14atg, which we have constructed from the observed broad-band
photometry ($uvw2,uvm2,uvw1,U,B,g,V,r,i$), adopting a distance modulus
of $34.85\pm0.1$\,mag according to NED\footnote{The NED estimate is
  based on a cosmology with $H_0=73$\,km\,s$^{-1}$\,Mpc$^{-1}$,
  $\Omega_\mathrm{m}=0.27$ and $\Omega_\Lambda = 0.73$, and a redshift
  of $z=0.021405$} and a Galactic extinction of
$A_B^\mathrm{gal} = 0.03$\,mag along the line of sight towards
IC~831. Looking at the \ions{Na}{i}\,D doublet in the spectra of
iPTF14atg, the host extinction is negligible and we adopt
$A_B^\mathrm{host}=0.00\pm0.02$\,mag. Errors on the bolometric flux
are dominated by the uncertainties on the distance modulus and the
extinction, and are indicated by a typical error bar in
Figure~~\ref{fig:bolo}. For the peak luminosity of iPTF14atg we find a
value of $\log\,L_\mathrm{peak}=42.58\pm0.04$ (log\,erg\,s$^{-1}$).

\section{Models}
\label{sec:models}

\subsection{Chandrasekhar-mass models}
\label{sec:mchmodels}

Owing to the early-time UV pulse, \citet{cao2015a} favour a SD origin
for iPTF14atg.  The most common explosion scenario for the SD
progenitor channel, is an explosion at the Chandrasekhar-mass
\citep[e.g.][]{wang2012b}. This scenario comes in several flavours
depending on the mode of flame propagation. A prompt detonation
produces almost pure \nuc{56}{Ni} ejecta \citep{arnett1969a} which is
not consistent with the observed properties of SNe~Ia
\citep[e.g.][]{mazzali2007a}. 

Instead, the delayed detonation scenario, where the flame starts as a
subsonic deflagration and transitions to a supersonic detonation at a
later stage \citep{khokhlov1991a}, has become the standard model to
produce SNe~Ia in the Chandrasekhar-mass scenario
\citep[e.g.][]{kasen2009a}. \citet{sim2013a} have recently explored
the observable parameter space of delayed detonations from a sample of
state-of-the-art 3D explosion models \citep{seitenzahl2013a} and find
peak bolometric luminosities in the range
$\log\,L_\mathrm{peak} = 42.80 - 43.31$ (log\,erg\,s$^{-1}$). 
Given the wide range of ignition configurations covered by these
models, it seems unlikely that a delayed detonation can account for
iPTF14atg ($\log\,L_\mathrm{peak}=42.58\pm0.04$). Moreover, the
synthetic spectra of the models do not match iPTF14atg: line expansion
velocities are generally too large and absorption features are too
strong, particularly before peak.

In the Chandrasekhar-mass scenario, weaker (and fainter) explosions
can be obtained from pure deflagrations. \citet{fink2014a} present
hydrodynamic explosion simulations and synthetic observables for a set
of 3D deflagration models in Chandrasekhar-mass CO WDs, covering a
wide range of ignition strengths. iPTF14atg lies well in the range of
peak bolometric luminosities $\log\,L_\mathrm{peak}=42.06 - 42.86$
(log\,erg\,s$^{-1}$) covered by these deflagrations models. The
closest match in terms of peak bolometric luminosity is provided by
model N5def ($\log\,L_\mathrm{peak}=42.59$). With five ignition
kernels in a small solid angle, N5def burns only a small fraction of
the initial WD (see figure~2 of \citealt{kromer2013a}).  Since the
nuclear energy release is less than the binding energy of the initial
WD, only $0.37$\,\msun\ of material are ejected in the explosion
(kinetic energy $1.34\times10^{50}$\,erg) and a bound remnant of
$\sim1.03$\,\msun\ is left behind. Owing to the turbulent evolution of
the deflagration flame, the ejecta are well mixed with a composition
predominantly of iron-group elements (IGEs, 0.222\,\msun), unburned O
(0.060\,\msun), C (0.043\,\msun) and intermediate-mass elements (IMEs,
0.042\,\msun).  With a \nuc{56}{Ni} yield of 0.158\,\msun\, the ejecta
of N5def give rise to a faint transient in good agreement to
SN~2005hk, a proto-typical SN~Iax (see \citealt{kromer2013a} for a
detailed discussion of the properties and synthetic observables of
N5def).  Here, we explore whether an explosion similar to N5def can
explain the observable properties of iPTF14atg.

\begin{figure*}
  \includegraphics[width=0.9\linewidth]{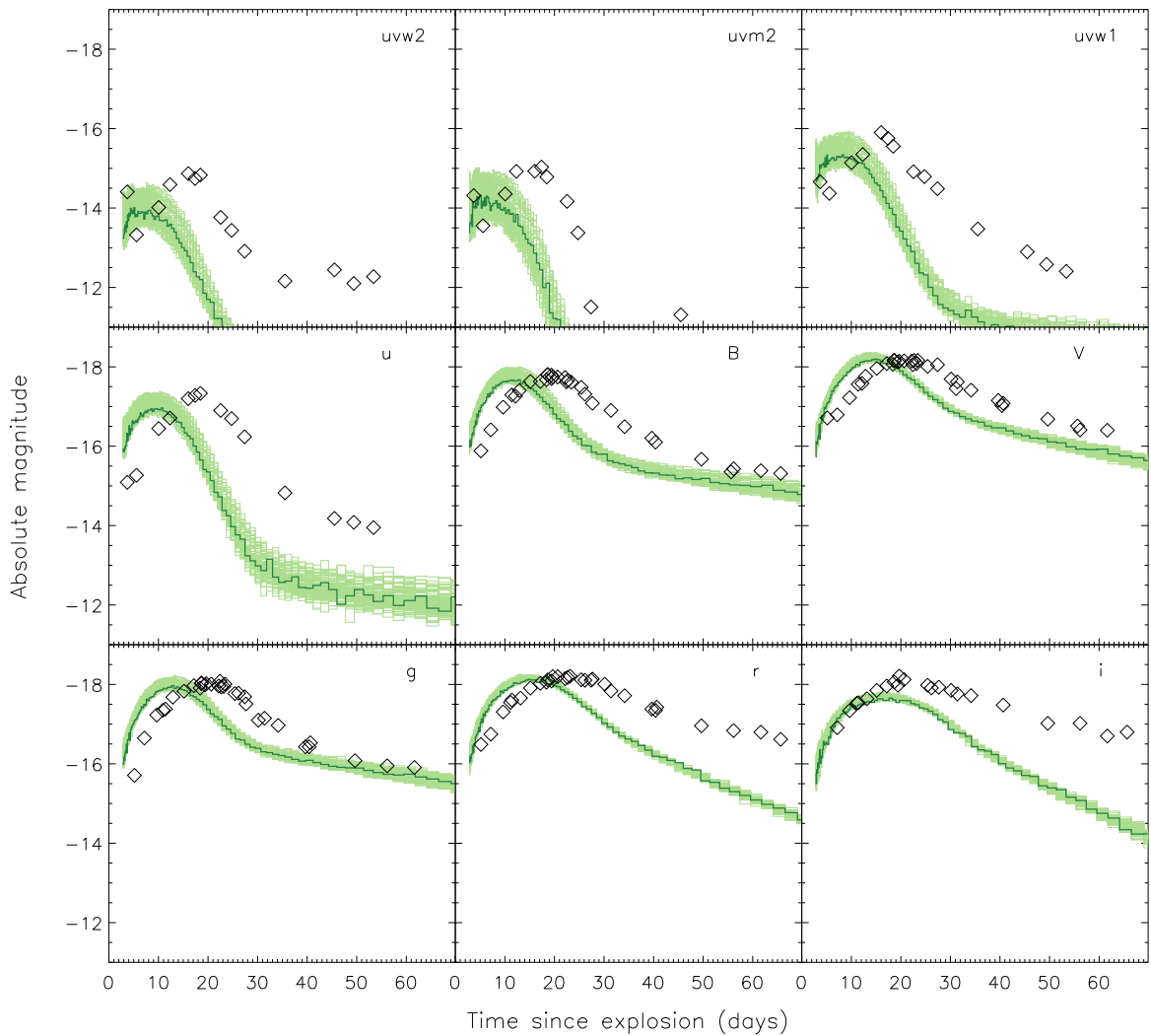}
  \caption{Broad-band synthetic light curves of N5def in various
    bands: \textit{Swift} UVOT $uvw2$, $uvm2$, $uvw1$, $u$, Bessell
    $B$, $V$ and SDSS $g$, $r$, $i$ (from top left to bottom right,
    respectively). UVOT and SDSS magnitudes are in the AB system,
    Bessell magnitudes in the Vega system. The region coloured in
    light green indicates the viewing angle diversity, while the dark
    green line shows a particular viewing angle (the same viewing
    angle has been selected for the different photometric bands and
    the spectral time series of the N5def model, presented in
    Figure~\ref{fig:spec_n5def}). The observed photometry of iPTF14atg
    is shown as black diamonds, assuming May 3.0 as the time of
    explosion \citep{cao2015a}, a distance modulus of 34.85\,mag for
    IC~831 and de-reddened for a Galactic extinction of
    $E(B-V)=0.11$.}
  \label{fig:lcs_n5def}
\end{figure*}

Figure \ref{fig:lcs_n5def} shows synthetic light curves of model N5def
along 100 different viewing angles in various photometric bands. The
diversity with viewing angle is strongest in the UV bands and
decreases at longer wavelengths. Overall the viewing angle sensitivity
is relatively modest since the turbulent deflagration produces
well-mixed ejecta. For comparison we overplot the observed photometry
of iPTF14atg as reported by \citet{cao2015a}. Although N5def provides
a good match regarding the peak brightness of iPTF14atg, the model's
time evolution is significantly too fast. In the $B$ band, for
example, the model peaks between 10.4 and 12.2\,d, depending on
viewing angle, while iPTF14atg peaked at 19.2\,d. This indicates that
the diffusion time is too short and the ejecta mass of N5def is
too low compared to iPTF14atg.

\begin{figure}
  \includegraphics[width=0.9\linewidth]{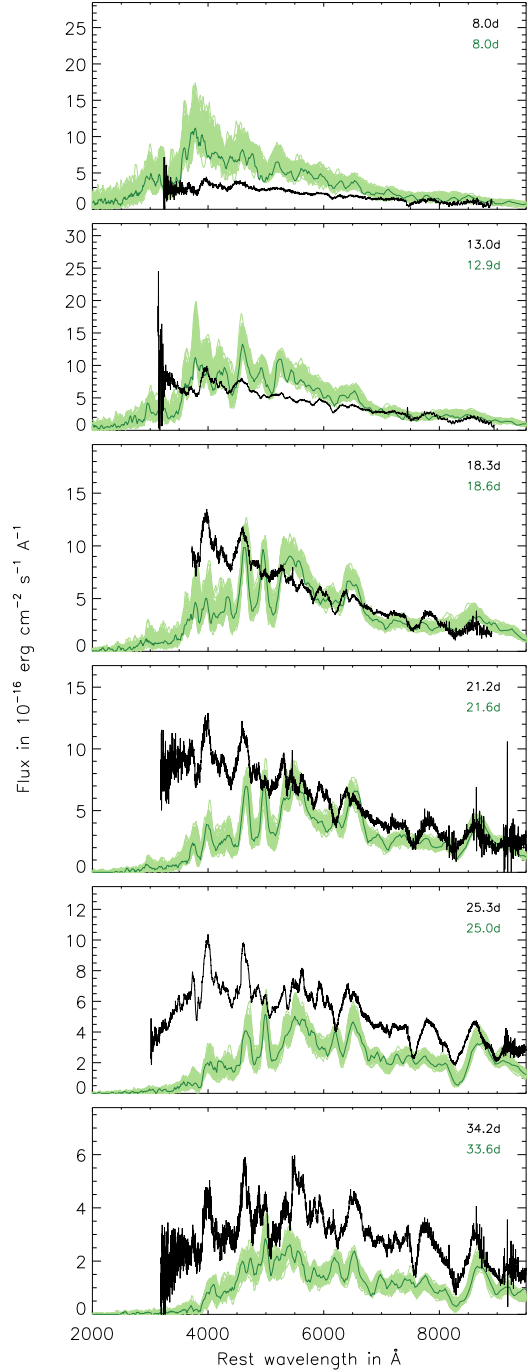}
  \caption{Snapshots of the spectral evolution of N5def for 6 epochs
    between 8 and 33.6\,d past explosion (the exact epochs are
    indicated in the individual panels). The dark green line shows
    spectra for a selected viewing angle (fixed for all epochs). To
    indicate the viewing angle diversity, we also show the spectra
    from 100 lines of sight (light green), which are equally distributed
    over the full solid angle. Observed spectra of iPTF14atg, which
    have been de-redshifted and de-reddened with $E(B-V)=0.011$ to
    account for the Galactic extinction towards IC~831, are shown in
    black for comparable epochs.}
  \label{fig:spec_n5def}
\end{figure}

Similar problems can be observed in the spectral evolution of the
N5def model. Figure \ref{fig:spec_n5def} shows synthetic spectra for
100 different viewing angles between 8 and 33.6\,d past explosion. As
for the light curves, the time evolution of the spectral time series
of N5def is significantly too fast compared to iPTF14atg. Moreover,
there are a number of prominent discrepancies between the spectral
features of the model and the data.  At all epochs, the model shows two
prominent emission features at $\sim$ 4500 and 5000\,\AA\ which are
not observed in iPTF14atg. Starting at 18.3\,d past explosion, the
model predicts too low flux levels blueward of $\sim$ 4500\,\AA\
indicating too much line blocking by the large fraction of IGEs in the
model ejecta.

More energetic deflagration models burn the full WD and lead to
larger ejecta masses \citep{fink2014a}. However, at the
same time they also produce significantly larger amounts of
\nuc{56}{Ni} in the ejecta, leading to peak bolometric luminosities
too bright for iPTF14atg [$\log\,L_\mathrm{peak}=42.77 - 42.86$
(log\,erg\,s$^{-1}$) for the models of \citealt{fink2014a}] and a
mismatch in the spectral evolution.

In summary, from our comparison between iPTF14atg and a wide variation
of Chandrasekhar-mass explosion models \citep{sim2013a,fink2014a}, we
find it unlikely that a Chandrasekhar-mass explosion can explain the
observational properties of iPTF14atg. Thus, we explore
alternative models in the following.

\subsection{Solar-metallicity merger model}

Given the likely classification of iPTF14atg as a 2002es-like SN~Ia,
violent mergers in WD binaries could be promising candidates to
explain the observed properties of iPTF14atg. For SN~2010lp, another
2002es-like SN~Ia, \citet{kromer2013b} reported remarkably good
agreement with the synthetic observables of a violent merger model.

In their particular model, \citet{kromer2013b} use the
smoothed-particle-hydrodynamics (SPH) code {\sc gadget}
\citep{springel2005a, pakmor2012b} to simulate the final inspiral of
two CO WDs with 0.90 and 0.76\,\msun, respectively (cf. figure~1 of
\citealt{kromer2013b}). Rapid accretion of material from the secondary
(less massive) WD onto the primary causes compressional heating
sufficient to ignite thermonuclear burning of C in the
simulation. Assuming that a detonation forms in the hottest cell,
\citet{kromer2013b} simulate the explosion dynamics with the
grid-based {\sc leafs} code \citep{reinecke2002b} and find that the
detonation completely disrupts the merged object within $\sim2$\,s,
ejecting $\sim1.6$\,\msun\footnote{When mapping the initial SPH
  simulation to the {\sc leafs} grid some mass is lost, since a number
  of SPH particles lie outside the simulated grid domain.} that stream
freely at $\sim100$\,s after the detonation (asymptotic kinetic energy
$1.1.\times10^{51}$\,erg). Since the detonation front propagates
faster in the high-density material of the primary WD, explosion ashes
from the primary can partially engulf the secondary before it is
disrupted. This leads to complex structures with pronounced
large-scale asymmetries and unburned O (from the secondary WD) in the
central parts of the ejecta (cf. figure~2 in \citealt{kromer2013b}).
The detailed nucleosynthetic yields of the explosive burning are
derived in a post-processing step with a 384-isotope nuclear network
\citep{travaglio2004a}, yielding a \nuc{56}{Ni} mass of
$0.18$\,\msun. To facilitate a comparison of the explosion model to SN
light curves and spectra, \citet{kromer2013b} obtain synthetic
observables with the multi-dimensional Monte Carlo radiative transfer
code {\sc artis} \citep{kromer2009a,sim2007a}. For more details on the
simulation setup and results see \citet{kromer2013b}.

\begin{figure*}
  \includegraphics[width=0.9\linewidth]{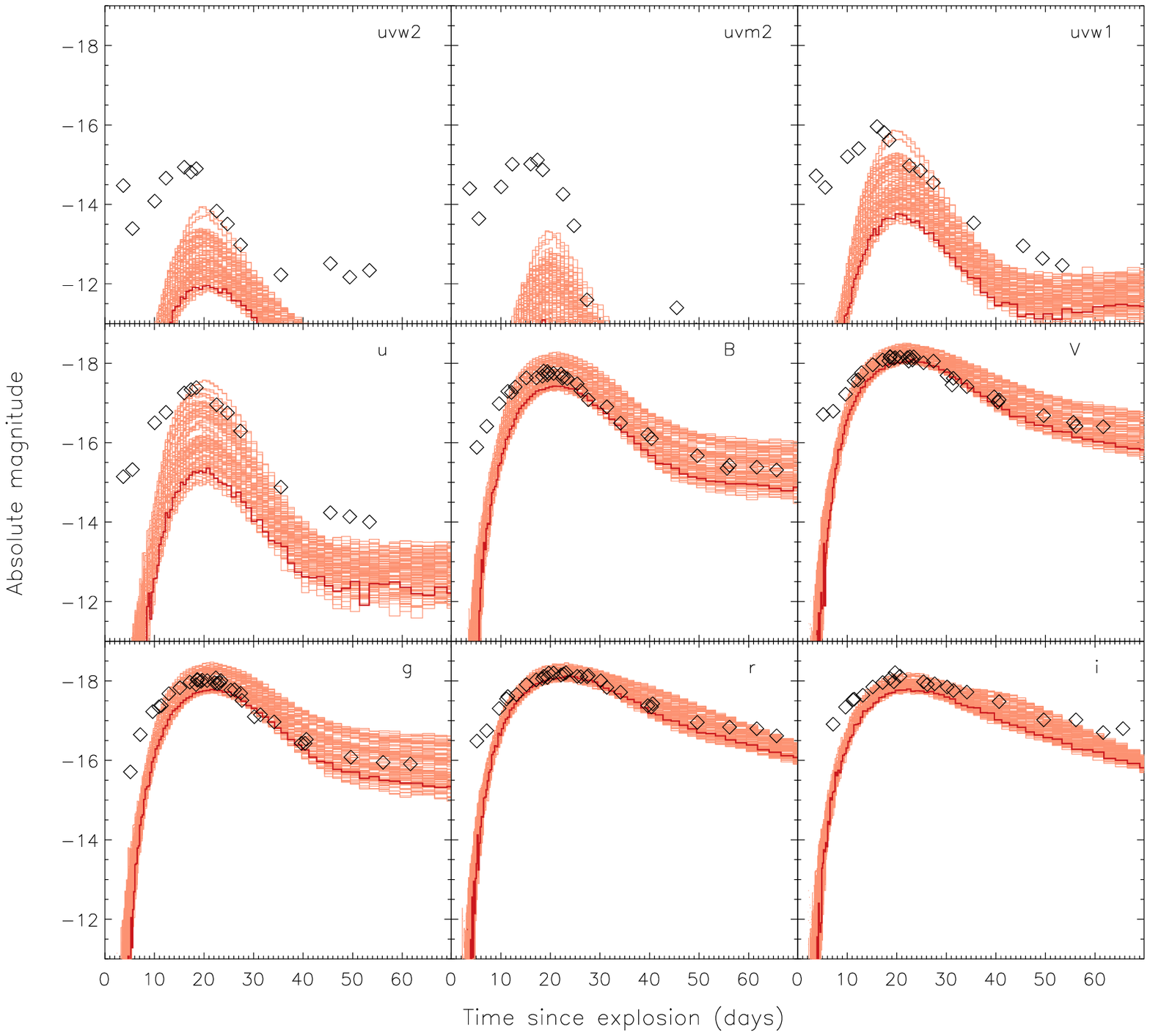}
  \caption{Broad-band synthetic light curves of the violent merger of
    two CO WDs ($0.90$+$0.76$\,\msun, \citealt{kromer2013b}) in
    various bands: \textit{Swift} UVOT $uvw2$, $uvm2$, $uvw1$, $u$,
    Bessell $B$, $V$ and SDSS $g$, $r$, $i$ (from top left to bottom
    right, respectively). UVOT and SDSS magnitudes are in the AB
    system, Bessell magnitudes in the Vega system. The region coloured
    in light red indicates the viewing angle diversity, while the dark
    red line shows a particular viewing angle [the same viewing angle
    has been selected for the different photometric bands and the
    spectral time series of the violent merger of \citet{kromer2013b},
    presented in the left panel of Figure~\ref{fig:spec_09076}]. The
    observed photometry of iPTF14atg is shown as black diamonds,
    assuming May 3.0 as the time of explosion \citep{cao2015a}, a
    distance modulus of 34.85\,mag for IC~831 and correcting for
    Galactic reddening.}
  \label{fig:lcs_09076_fullZsol}
\end{figure*}

Here, we take the model of \citet{kromer2013b} and compare it to the
observed properties of iPTF14atg. Figure \ref{fig:lcs_09076_fullZsol}
shows synthetic light curves of the model along 100 different viewing
angles in various photometric bands. Owing to the large-scale
asymmetries in the merger ejecta, the light curves show a prominent
viewing angle sensitivity. After 10\,d past explosion, the merger
model matches iPTF14atg reasonably well in $V$ and redder bands. At
earlier epochs and in the bluer bands the model shows a pronounced
flux deficit compared to iPTF14atg.

\begin{figure*}
  \includegraphics[width=0.45\linewidth]{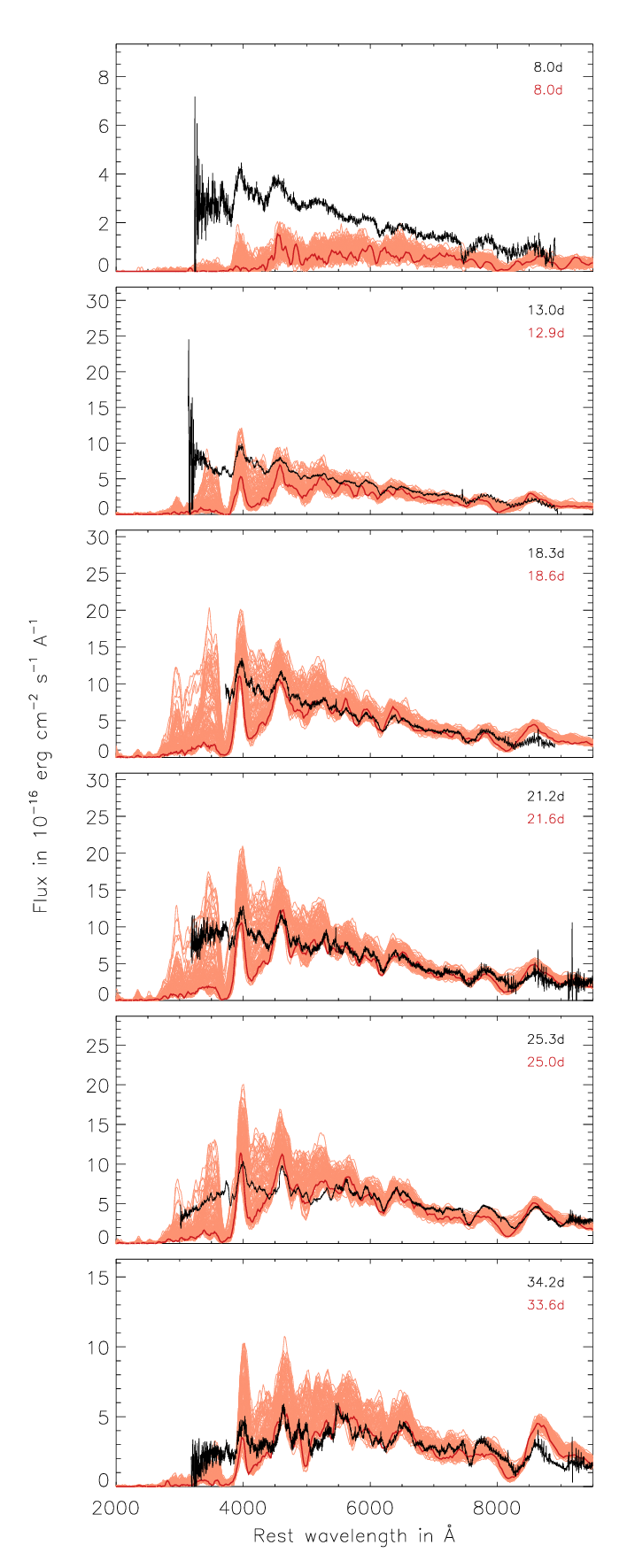}
  \includegraphics[width=0.45\linewidth]{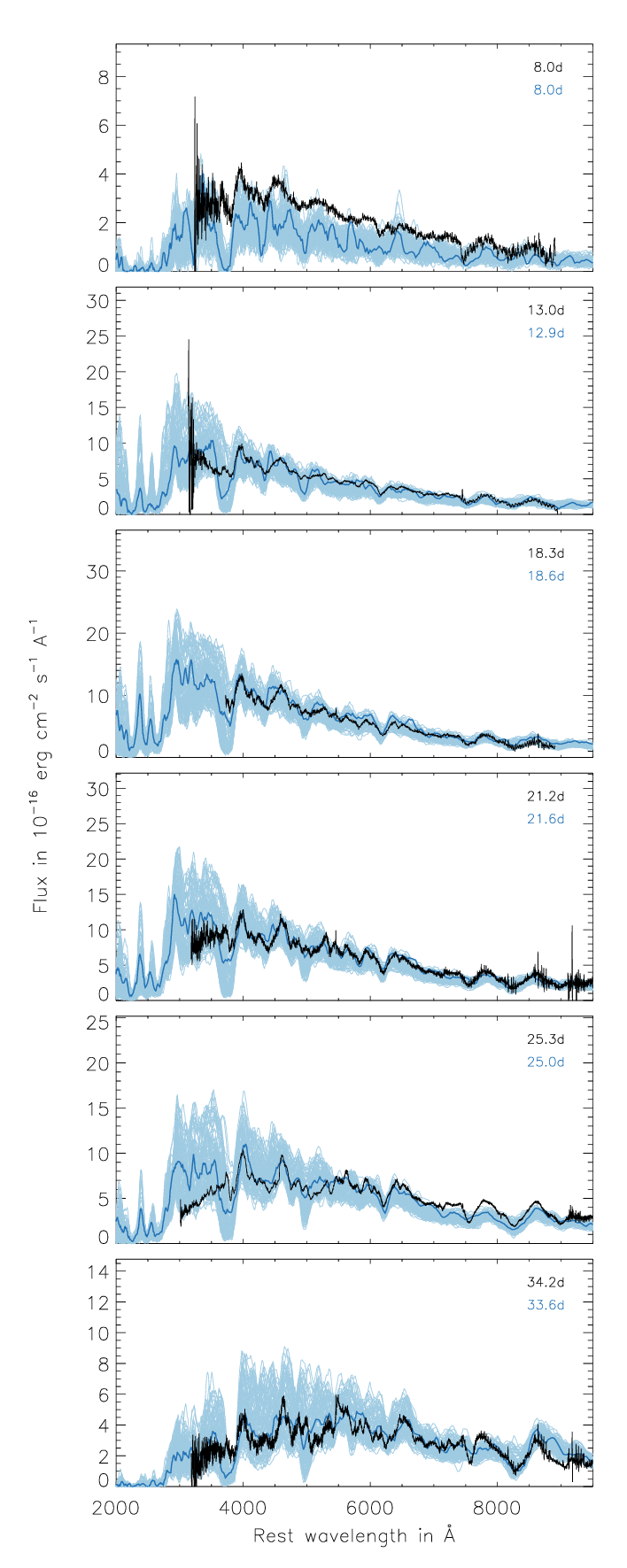}
  \caption{\textit{Left:} Synthetic spectra for the violent merger
    model of \citet{kromer2013b} where the progenitor WDs have been
    admixed with metals heavier than C and O in solar composition. The
    dark red line shows spectra for a selected viewing angle (fixed
    for all epochs). To indicate the viewing angle diversity, we also
    show the spectra from 100 lines of sight (light red), which are
    equally distributed over the full solid angle. Observed spectra of
    iPTF14atg (de-redshifted and de-reddened) are shown in black for
    comparable epochs. \textit{Right (in blue colour scheme):} Same as
    left hand panel but for a modified merger model from a progenitor
    system with reduced metallicity ($Z=0.01Z_\odot$). }
  \label{fig:spec_09076}
\end{figure*}

The same behaviour can be seen for the spectral evolution
(Figure~\ref{fig:spec_09076}, left panel). At 8\,d the model flux is
significantly lower than the observed spectrum of iPTF14atg. From 12.9
to 33.6\,d the overall spectral energy distribution (SED) of the model
is in good agreement with the observed SED of iPTF14atg. Small
differences in the SED become visible where the observed spectra
extend to the UV, indicating a deficit of UV photons in the model.

Taking a closer look, the merger spectra also reproduce the key
spectral features of iPTF14atg. A striking difference, however, is
visible for the \ions{Ti}{ii} dominated region between 4000 and
4400\,\AA\ which shows a strong absorption trough in the model spectra
that is not present in iPTF14atg. A similar behaviour can be observed
for the \ions{Ca}{ii} features. Both the H\&K doublet and the NIR
triplet are far too strong in the model, particularly at
early epochs. The latter problem might be related to the composition
of the progenitor WDs in the model of \citet{kromer2013b}. Assuming
WDs from main-sequence progenitors with solar metallicity ($Z_\odot$),
\citet{kromer2013b} adopt an initial WD composition of 50\% O and
48.29\% C (by mass). The remaining 1.71\% are distributed according to
the solar abundances of \citet{asplund2009a} for all elements but H
and He (primordial C, N, and O were converted to \nuc{22}{Ne} to
account for core He-burning).  This introduces a significant
contribution of Ca and Ti in the unburned outer ejecta, which may be
responsible for the deep and broad Ca and Ti features.

\subsection{Reducing the metallicity of the merger model}

For a low-metallicity progenitor system the unburned outer layers will
have lower Ca and IGE contents. This might help to alleviate some of
the shortcomings discussed in the previous paragraph.  To explore the
influence of this effect, we have repeated the postprocessing and
radiative transfer simulations for a lower metallicity progenitor
system of our $0.90$+$0.76$\,\msun\ merger. Specifically, we have
assumed $Z=0.01Z_\odot$ for the progenitor WDs. The spectral time
series of this new calculation is shown in the right hand panel of
Figure~\ref{fig:spec_09076}. Compared to the original model, the Ca
features are significantly weaker, which brings the model in better
agreement with the observed features of iPTF14atg. Remarkably, the new
model also agrees almost perfectly with iPTF14atg in the \ions{Ti}{ii}
dominated region between 4000 and 4400\AA. The outer ejecta layers of
our modified merger model contain much less Ti than those of the
original model of \citet{kromer2013b} that started from solar
metallicity progenitors.  Consequently, the \ions{Ti}{ii} absorption
is significantly lower in our new model, leading to much better
agreement with iPTF14atg.

Moreover, we find a notably increased UV flux for the new model while
the flux level between 4500 and 7000\,\AA, is slightly reduced.  This
improves the agreement of the model SED with the data.  In the
original model the unburned outer ejecta layers are polluted with IGEs
in solar composition. This leads to a strong suppression of UV flux by
line blocking which is then redistributed to optical wavelengths,
where the radiation escapes from the ejecta. In the new model the IGE
content of the unburned layers is reduced by a factor 100, leading to
significantly less line blocking. As a consequence the new model also
evolves slightly faster. We find a $B$-band peak time of 19.8\,d which
is in good agreement with the rise time of 19.2\,d of iPTF14atg [the
original merger model of \citet{kromer2013b} has a rise time of
21.3\,d, for comparison]. This also leads to an increased flux at
early epochs and brings the model in closer agreement with the 8\,d
spectrum, although the model flux is still too low for the majority of
viewing angles.

\begin{figure*}
  \includegraphics[width=0.9\linewidth]{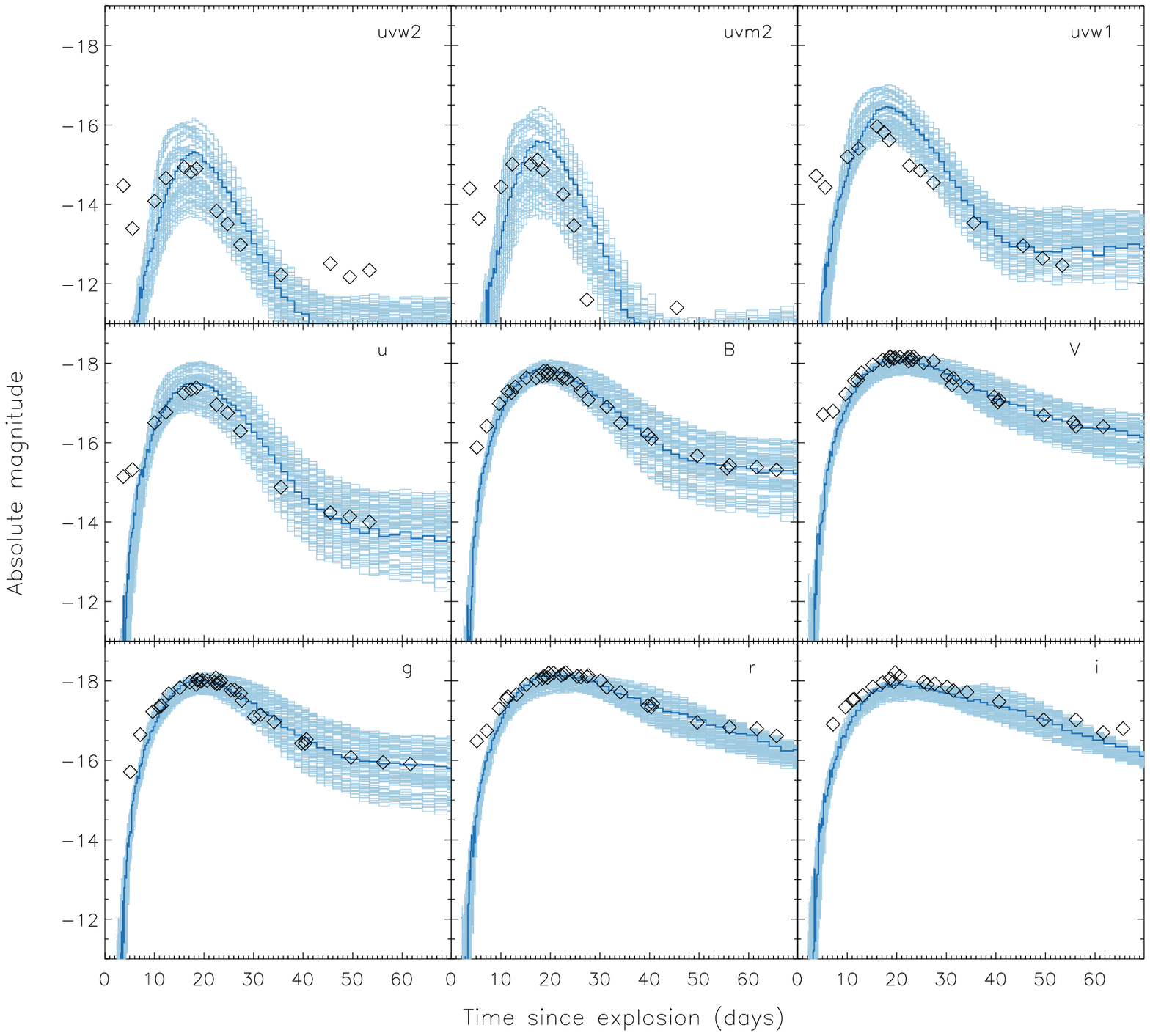}
  \caption{Broad-band synthetic light curves of our new
    $0.90$+$0.76$\,\msun\ violent merger model with $Z=0.01Z_\odot$:
    \textit{Swift} UVOT $uvw2$, $uvm2$, $uvw1$, $u$, Bessell $B$, $V$
    and SDSS $g$, $r$, $i$ (from top left to bottom right,
    respectively). UVOT and SDSS magnitudes are in the AB system,
    Bessell magnitudes in the Vega system. The region coloured in
    light blue indicates the viewing angle diversity, while the dark
    blue line shows a particular viewing angle (the same viewing angle
    has been selected for the different photometric bands and the
    spectral time series of our $Z=0.01Z_\odot$ violent merger,
    presented in the right panel of Figure~\ref{fig:spec_09076}). The
    observed photometry of iPTF14atg is shown as black diamonds,
    assuming May 3.0 as the time of explosion \citep{cao2015a}, a
    distance modulus of 34.85\,mag for IC~831 and correcting for
    Galactic reddening.}
  \label{fig:lcs_09076_zeroZ}
\end{figure*}

Figure~\ref{fig:lcs_09076_zeroZ} shows synthetic light curves for the
new model. After about 10\,d past explosion the agreement between the
new model and iPTF14atg in the optical bands is excellent. In the
\textit{Swift} UV filters ($uvw2$,$uvm2$,$uvw1$), the reduced line
blocking leads to significantly increased peak flux and much better
agreement with iPTF14atg. While the original model showed a clear flux
deficit in these bands (compared to the data), some lines-of-sight of
the new model have a UV flux level comparable to iPTF14atg. In fact,
the majority of viewing angles has higher UV flux than iPTF14atg. This
leaves room for a progenitor system with sub-solar metallicity for
iPTF14atg.  Constraining the chemical composition of the progenitor
system more precisely, would require a large set of models for
different metallicities. But even then the large viewing-angle
sensitivity in the UV will make it difficult to determine an exact
value for the progenitor composition. In addition, other parameters
like the initial masses of the WDs or the time of ignition will also
affect the observational display, making the potential parameter space
even larger. We thus do not aim to obtain a perfect fit with our new
$Z=0.01Z_\odot$ model, but stress that a violent merger with low
progenitor metallicity can provide an excellent match to iPTF14atg.

We also note that our merger model contains a significant fraction of
O in the central parts of the ejecta, which originates from the
secondary WD and is at too low initial density to be burned within the
explosion (cf. figure~2 in \citealt{kromer2013b}). This could lead to
[\ions{O}{i}] emission as observed in iPTF14atg, given that the
excitation and ionization conditions are appropriate. Unfortunately,
we can currently not simulate this directly, since our 3D radiative
transfer code {\sc artis} does not account for the necessary
non-thermal excitation and ionization processes. One-dimensional
models with a more sophisticated excitation treatment are only of
limited use. Mapping the highly asymmetric ejecta of our merger model
to 1D leads to artificial mixing of chemical species, which introduces
a multitude of additional complications.  A detailed investigation of
the late-time spectrum must therefore be postponed to future work.

\section{Discussion}
\label{sec:discussion}

Our models indicate that it will be difficult to explain iPTF14atg in
the context of Chandrasekhar-mass explosions. Lowering the metallicity
of the progenitor system will not significantly affect the observables
in this case, since burning ashes are mixed to the outer ejecta during
the turbulent deflagration phase \citep[e.g.][]{seitenzahl2013a}.
Consequently, the IGE mass fraction is well above the solar
composition in most of the ejecta (for pure deflagration models, like
N5def presented in Section~\ref{sec:mchmodels}, over the full velocity
range of the ejecta). Thus, an 0.1\% contribution of progenitor IGEs
for a model with solar metallicity does not have a strong impact on
the observables of Chandrasekhar-mass explosions. This is only the
case, if there are large parts of the ejecta that do not contain
freshly synthesized IGEs like in our merger model.

An alternative SD progenitor option could be a double detonation in a
He-accreting sub-Chandrasekhar-mass WD.  If accreting from a
non-degenerate He star donor, such a system belongs to the SD
progenitor class. A number of simulations have shown that double
detonations can lead to faint SNe~Ia \citep[e.g.][]{hoeflich1996a,
  fink2010a, woosley2011a}.  However, there are several potential
problems. (i) It has been predicted that the ashes of the first
detonation in the He shell should leave characteristic imprints around
maximum light \citep{kromer2010a}, which are not observed in the case
of iPTF14atg. (ii) Having an explosion below the Chandrasekhar-mass
will lead to low ejecta mass and short diffusion time -- in contrast
to the slow evolution observed for iPTF14atg. In fact,
sub-Chandrasekhar double detonations have been proposed to explain the
rapidly evolving subluminous 1991bg-like SNe~Ia, owing to their fast
light curve evolution \citep{stritzinger2006a,scalzo2014a}. (iii) With
a He-accreting double detonation model it will be difficult to explain
the 6300\,\AA\ feature in the late-time spectrum.  [\ions{O}{i}]
emission seems impossible since no O will be left in the central
ejecta. (iv) In addition, for the case of a double detonation in a SD
progenitor system, i.e. a sub-Chandrasekhar-mass WD accreting from a
non-degenerate He star donor, the theoretical delay-time distribution
is strongly peaked at short times ($< 500$\,Myr,
\citealt{ruiter2011a}). This makes such systems unlikely progenitors
for 2002es-like SNe, which have been observed preferentially (but not
exclusively) in massive early-type host galaxies \citep{white2015a}.

Taking all together, from the models we have at hand a SD progenitor
of iPTF14atg seems unlikely.  In contrast, the merger model presented
in the previous section shows excellent agreement with the observed
properties of iPTF14atg after about 10\,d past explosion. There are,
however, some discrepancies at the earliest epochs. In particular, our
model cannot naturally explain the early UV pulse in the
\textit{Swift} observations (compare e.g.\
Figure~{\ref{fig:lcs_09076_zeroZ}). In the following, we discuss
possibilities how to account for the early-time UV pulse in a merger
scenario.

\subsection{CSM interaction}

In some core-collapse explosions of massive stars a similar behaviour
to the early light curve of iPTF14atg is observed
\citep[e.g.][]{gezari2008a}.  This is attributed to cooling of shock
heated ejecta immediately after the shock breakout
\citep{rabinak2011a}.  For exploding WDs the radius is so small that
the time scales for shock breakout are very short and the additional
luminosity is too faint to be detected \citep[e.g.][]{piro2010a,
  rabinak2012a}.  However, it has been argued that shock interaction
with an extended circum-stellar medium (CSM), which happens on longer
time scales, could lead to additional early-time optical/UV
emission \citep[e.g.][]{raskin2013b,levanon2015a}. 

\citet{cao2015a} have ruled out such a scenario for iPTF14atg.
However, they have only investigated the special case of an extended
optically thin CSM in a spherically symmetric configuration. The
potential parameter space for the CSM configuration of DD mergers is
however larger and a spherically symmetric configuration seems highly
unlikely for systems that explode on the dynamical or viscous time
scale.

\citet{raskin2013b} discuss the signatures of CSM interaction in WD
mergers for a variety of CSM configurations with a focus on tidal
tails (see also \citealt{levanon2015a}). Depending on the lag time
between the time of the tidal tail ejection and the time when the
merged system explodes, they find a wide range of observables.  For
lag times on the order of the viscous time scale of the merger, the
CSM is not very extended ($r \lesssim 10^{13}$\,cm). For this case
\citet{raskin2013b} predict a soft ($\sim$100\,eV) shock breakout
signal from tidal tail interaction that lasts for a few minutes and is
followed by bright optical/UV cooling emission
($L_\mathrm{bol} = 10^{42}-10^{43}$\,erg\,s$^{-1}$). This is fairly
similar to the early-time UV luminosity
($L_\mathrm{UV} = 3\times10^{41}$\,erg\,s$^{-1}$) that
\citet{cao2015a} reported for iPTF14atg. However, \citet{raskin2013b}
also note that, owing to the low mass of the tidal tails, the duration
of the optical/UV cooling emission should be fairly short. They
estimate a duration of about half a day. This is too short to explain
the initial UV pulse of iPTF14atg, which lasted for several days.

Sophisticated radiation hydrodynamics simulations, exploring a wider
range of CSM configurations for merger models from various initial
parameters, will be required to investigate the observable display of
the CSM interaction scenario in more detail and evaluate its potential
to explain the early-time luminosity of iPTF14atg. However, it is
interesting to note that \cite{piro2015a} have recently shown that CSM
interaction can -- at least in the optical -- lead to similar effects
in the early-time light curves as for the case of interactions between
the SN ejecta and a non-degenerate companion star \citep{kasen2010a}.

\subsection{Surface radioactivity}

Another possibility to explain the early-time UV pulse could be
radioactive material close to the ejecta surface. This is
characteristic for double detonations in He accreting systems
\citep[e.g.][]{fink2010a,woosley2011b}. Recently, high resolution
simulations of WD mergers by \citet{pakmor2013a} and
\citet{tanikawa2015a} have shown that even a thin He layer, which is
expected on most CO WDs according to stellar evolution calculations,
is sufficient to lead to a He shell detonation (the resolution of the
merger model presented here is not high enough to resolve such a He
layer). Both \citet{pakmor2013a} and \citet{tanikawa2015a} find that
this initial He detonation can trigger a secondary detonation in the
more massive component of the merger by converging shock fronts
(He-ignited violent merger scenario). In this case the observational
display of the explosion would be rather similar to a
sub-Chandrasekhar-mass double detonation with all the problems
discussed above.

However, \citet{tanikawa2015a} note that the triggering of the
secondary detonation could also fail e.g.\ due to inhomogeneities in
the He shell or instabilities in the burning. In this case the system
survives the He shell detonation and a ``classical'' carbon-ignited
violent merger explosion might occur at a later epoch, given that
sufficiently high temperatures and densities are reached. The
observational display would then be very similar to a ``classical''
carbon-ignited violent merger explosion as presented in this paper,
with the additional complication of ashes from the initial He shell
detonation at the surface of the ejecta.

\citet{cao2015a} have shown that a shell of 0.01\,\msun\ of
\nuc{56}{Ni} would be required to explain the early time UV flux of
iPTF14atg, which is comparable to theoretical predictions in double
detonation models \citep[e.g.][]{fink2010a,woosley2011b}. However,
\citet{cao2015a} disfavour surface radioactivity as the source for the
early time UV flux. They argue that flux redistribution to the
optical/NIR would suppress the UV flux \citep{kromer2010a}. However,
the discussion in \citet{kromer2010a} refers to effects around the
peak of the optical light curves. At these epochs iron-group elements
in the He shell ashes act mainly as an opacity source and are very
efficient in redistributing blue photons that originate from
\nuc{56}{Ni} in the centre of the ejecta (ashes from the core
detonation) to redder wavelengths when propagating towards the surface
of the ejecta. At early times the situation is
different. There, the densities are still high so that local
deposition of radioactive energy in the shell material dominates,
thereby heating the shell and leading to a strong UV flux.

As outlined above, the critical point for a scenario with a He surface
detonation is to have sufficient radioactive material in the outer
layers to explain the early UV flash, but at the same time not too
much iron-group elements to avoid strong flux redistribution at
optical peak. In double detonations with massive He shells that are
ignited by a thermal instability this seems difficult, since the shell
masses are fairly large \citep[e.g.][]{kromer2010a, woosley2011b}.
However, in mergers it is possible to ignite much smaller He shell
masses dynamically \citep[e.g.][]{guillochon2010a,
  pakmor2013a}. Whether such models produce enough surface
radioactivity to account for the early-time UV luminosity of
iPTF14atg, while not causing strong observable imprints around maximum
light, requires a dedicated study with high-resolution simulations.

Alternatively, geometry effects could play a role. If, for example,
IGEs and radioactive isotopes are confined to a narrow ring at the
surface of the ejecta, an observer looking down a line of sight
perpendicular to this ring could see additional luminosity from the
surface radioactivity at early epochs, while later on the ring
material will not provide an opacity source for radiation originating
from the core of the ejecta. Such a geometry has been proposed by
\citet{diehl2014a} to explain early-time $\gamma$-ray observations
of SN~2014J obtained with the \textit{Integral} satellite.

\section{Conclusions}
\label{sec:conclusions}

We have compared synthetic observables of state-of-the-art
multi-dimensional explosion models to the light curves and spectra of
the subluminous 2002es-like SN iPTF14atg.  While the detection of an
early-time UV spike in iPTF14atg was interpreted as evidence for a SD
progenitor system in a previous analysis by \citet{cao2015a}, we have
difficulties explaining the spectral evolution of iPTF14atg within the
SD progenitor channel from our models. Specifically, we find that the
failed deflagration of a Chandrasekhar-mass WD, which could reproduce
the observed luminosity of iPTF14atg quite well, has too fast evolving
light curves compared to iPTF14atg and cannot account for the observed
spectral features.

In contrast, we find reasonable agreement between the spectral
evolution of a violent merger of two CO WDs with 0.90 and 0.76\,\msun,
respectively, which had previously been suggested as progenitor for
the 2002es-like SN~2010lp \citep{kromer2013b}. Minor differences in
the spectral features between this model, for which solar metallicity
main-sequence progenitors were assumed, and iPTF14atg point at a
mismatch in progenitor metallicity. Repeating the merger simulation
for a progenitor system with sub-solar metallicity ($Z=0.01Z_\odot$),
we find remarkably good agreement of the synthetic light curves and
spectra with iPTF14atg for epochs after 10\,d past explosion,
suggesting a low-metallicity progenitor system for iPTF14atg.

At earlier epochs the flux level is slightly too low and our model
cannot explain the early UV spike naturally. We argue that an
additional energy source is required to explain the observed flux at
these early epochs and discuss cooling of a shock heated CSM and
surface radioactivity from a double detonation as possibilities to
account for the missing flux at early epochs with DD progenitors. In
the light of recent studies \citep[][]{raskin2013b, levanon2015a}, CSM
interaction seems to be the most promising solution. Sophisticated
radiation hydrodynamics simulations will be required to address this
in more detail. Violent merger scenarios where an initial detonation
of a He shell leads to the production of radioactive isotopes close to
the surface of the ejecta, may also be a possible solution, if the
peculiar surface composition will not leave a strong observable
imprint around maximum light. A new generation of high-resolution WD
merger models will be required, to investigate whether this is
possible.

\section*{Acknowledgements}

We gratefully acknowledge support from the Knut and Alice Wallenberg
Foundation. The Oskar Klein Centre is funded by the Swedish Research
Council. The work of RP and FKR is supported by the Klaus Tschira
Foundation. RP also acknowledges support by the European Research
Council under ERC-StG grant EXAGAL-308037. ST is supported by the
Deutsche Forschungs\-gemeinschaft via the Transregional Collaborative
Research Center TRR 33 `The Dark Universe'. AG and RA acknowledge
support from the Swedish Research Council and the Swedish National
Space Board. IRS was supported by the Australian Research Council
Laureate Grant FL0992131.

The authors gratefully acknowledge the Gauss Centre for Supercomputing
(GCS) for providing computing time through the John von Neumann
Institute for Computing (NIC) on the GCS share of the supercomputer
JUQUEEN \citep{stephan2015a} at J\"ulich Supercomputing Centre
(JSC). GCS is the alliance of the three national supercomputing
centres HLRS (Universit\"at Stuttgart), JSC (Forschungszentrum
J\"ulich), and LRZ (Bayerische Akademie der Wissenschaften), funded by
the German Federal Ministry of Education and Research (BMBF) and the
German State Ministries for Research of Baden-W\"urttemberg (MWK),
Bayern (StMWFK) and Nordrhein-Westfalen (MIWF).

\label{lastpage}
\end{document}